\documentclass[a4paper, pra, aps, nopacs, twocolumn, superscriptaddress]{revtex4-2}

\usepackage[utf8]{inputenc}  
\usepackage[british]{babel}  
\usepackage{graphicx} 

\usepackage{relsize}

\usepackage{epsfig}

\usepackage{pgfplots}

\usepackage{dcolumn}   
\usepackage{bm}        
\usepackage{amssymb}   
\usepackage{amsmath}

\usepackage{nicematrix}
\usepackage{mathtools}
\usepackage{tikz}
\usepackage{mdframed}
\usepackage{theorem}

\usepackage[bookmarks=false]{hyperref}
\hypersetup{colorlinks=true,citecolor=blue,linkcolor=red,
urlcolor=blue,pdfstartview=FitH,bookmarksopen=true}

\DeclareMathOperator{\Tr}{Tr}

\usepackage{float}

\newcommand{\ket}[1]{\ensuremath{|#1\rangle}}

\newcommand{\bra}[1]{\ensuremath{\langle#1|}}
\newcommand{\BraKet}[2]{\ensuremath{\langle #1|#2\rangle}}

\newcommand{\abs}[1]{|#1|}

\newcommand{\ba}{\begin{eqnarray}}
\newcommand{\ea}{\end{eqnarray}}
\newcommand{\ban}{\begin{eqnarray*}}
\newcommand{\ean}{\end{eqnarray*}}
\newcommand{\be}{\begin{equation}}

\newcommand{\ee}{\end{equation}}
\newcommand{\rme}{\ensuremath{\mathrm{e}}}
\newcommand{\rmi}{\ensuremath{\mathrm{i}}}

\begin{document}
\nonfrenchspacing
\title{Characterizing generalized axisymmetric quantum 
states in $d\times d$ systems}

\author{ Marcel Seelbach Benkner}
\affiliation{Naturwissenschaftlich-Technische 
Fakultät, Universität Siegen, Walter-Flex-Straße 3, 57068 Siegen, Germany}

\author{Jens Siewert}
\affiliation{Departamento de Qu\'{i}mica Física, 
Universidad del Pa\'{i}s Vasco UPV/EHU, 
48080 Bilbao, Biscay, Spain}
\affiliation{Ikerbasque Basque Foundation for Science, 48013 Bilbao, Biscay, Spain}
\author{Otfried Gühne}
\affiliation{Naturwissenschaftlich-Technische 
Fakultät, Universität Siegen, Walter-Flex-Straße 3, 57068 Siegen, Germany}
\author{Gael Sentís}
\affiliation{Naturwissenschaftlich-Technische 
Fakultät, Universität Siegen, Walter-Flex-Straße 3, 57068 Siegen, Germany}
\affiliation{Física Teòrica: Informació i Fenòmens Quàntics, Departament de Física,
Universitat Autònoma de Barcelona, 08193 Bellatera, Barcelona, Spain}

\date{\today}  

\begin{abstract}
We introduce a family of highly symmetric bipartite quantum states in arbitrary dimensions. It consists of all 
states that are invariant under local phase rotations
and local cyclic permutations of the basis. We solve the separability problem for a subspace of these states and 
show that a sizable part of the family is bound entangled.
We also calculate some of the Schmidt numbers for the family 
in $d = 3$, thereby characterizing the dimensionality of entanglement. Our results allow us to estimate entanglement
properties of arbitrary states, as general states can be symmetrized to the considered family by local operations.
\end{abstract}
\maketitle

\section{Introduction}
Entanglement remains as one of the most significant features that distinguishes quantum theory from a classical description of our world. It also underpins a wide range of quantum information processing tasks, such as quantum key distribution~\cite{Curty2004}, quantum communication~\cite{HorodeckiRMP2009}, and quantum sensing~\cite{Pezze2009}, where it acts as a resource for increased performance. Detecting, characterizing, and quantifying the various forms of entanglement present in quantum states is therefore of high practical relevance as well as of fundamental interest~\cite{Dada2011}, shedding light onto the principles that build our physical world through the structure of possible correlations between multiple systems.

While detecting entanglement in pure states shared among two parties is trivial, there is no infallible method to decide whether an arbitrary (noisy) bipartite state is entangled~\cite{entanglement}. Beyond detecting its presence, it is also important to identify its character. For instance, entanglement can be of high dimension, which provides additional advantages such as noise resistance in entanglement distribution protocols~\cite{Ecker2019} or increased capacity of quantum communication~\cite{Mirhosseini2015}.
Consequently, considerable effort is devoted to creating such high-dimensional entangled states in experiments~\cite{erhard2020advances}.

Entanglement may also be bound, that is, a form of entanglement that cannot be distilled into its purest form of maximal dimensionality 
by local operations and classical communication~\cite{PhysRevLett.80.5239}, which is the resource required by most standard applications in quantum information. The observation that known bound entangled states typically lie in close proximity to the separable states has led to the belief that bound entanglement is a ``weak'' form of entanglement, thus useless for quantum information processing. Surprisingly, though, it has been established that bound entangled states may still serve as a resource for certain tasks, e.g., for quantum key distribution and entanglement activation~\cite{Horodecki2005,PhysRevLett.82.1056,Masanes2006}. The exact relation between 
distillability and dimensionality of entanglement is an actively researched open question~\cite{schmidt,Szarek2010,YANG2016,Chen2017,Huber2018}.

In this paper, we explore these two features of entanglement in
families of states that obey certain symmetries.
Characterizing the entanglement properties of symmetric families has two advantages: First, the reduced number of free parameters makes an otherwise daunting task easier, and second, the symmetries that generate the family provide a simple method to obtain lower bounds on the amount of entanglement (measured either quantitatively or by its dimension) of arbitrary mixed states, by ``twirling'' them into the family~\cite{Sentis2016d}. 
In addition, fully characterizing the set of separable states in symmetric families can even help in establishing results on entanglement theory that go beyond that subset of states. A recent example is the superadditivity of genuine multipartite entanglement~\cite{Palazuelos2022}, the proof of which relies on having a full characterization of separable isotropic states~\cite{IsotropicStates}.
To detect the undistillability of entangled states, we resort to the positive partial transpose (PPT) criterion, as entangled states which meet the PPT criterion are known to be bound entangled~\cite{Horodecki1996,PhysRevLett.77.1413}. We also look at the Schmidt numbers of the states in the family, a paradigmatic measure of entanglement dimensionality~\cite{terhal2000schmidt}.

With these techniques at hand, we study a highly symmetric family of bipartite mixed states with local dimension $d$, introduced in Refs.~\cite{Main,Baumgartner2006,Bertlmann2007,Bertlmann2008,Bertlmann2008a,Bertlmann2009}, fully characterizing the set of separable states in a subset of this family, and thereby providing a complete characterization of all PPT entangled states for arbitrary $d$. We then use fidelity-based Schmidt number witnesses, together with a combination of analytical and numerical methods, to further refine the characterization of the sets of states with varying Schmidt number. We show explicit results for $d=3$ and $d=4$, although our methodology is extensible to higher local dimensions. Our results comprise an exceptional example of a rich family of states arising from symmetries where their entanglement properties can be described to a very high degree.

\section{Basic notions: entanglement \& symmetry}

Throughout this paper we denote by $\varrho$ a density matrix that describes a quantum state, which is shared between the two local parties Alice and Bob. We only consider scenarios where the dimensions of the local Hilbert spaces  are equal and denote this dimension by $d$. 
One central question in quantum information theory is whether or not a quantum state $\varrho$ is entangled. It is well known that for pure quantum states one can calculate the Schmidt rank to answer this question. Pure quantum states are separable if and only if they have Schmidt rank $1$ and thus can be written as $\ket{\psi}= \ket{\psi_A}\otimes \ket{\psi_B}$. Otherwise they are entangled. 
\\\\

\paragraph{Schmidt numbers:} In Ref. \cite{terhal2000schmidt} the definition of the Schmidt rank is extended to mixed states and is called the Schmidt number. To determine if the Schmidt number of a state $\varrho$, is smaller than or equal to $K$, one asks for a decomposition of $\varrho$ into pure states $\ket{\psi_i}$ where all Schmidt ranks $ \mathcal{S}( \ket{\psi_i})$ are smaller than or equal to $K$. To obtain the Schmidt number of $\varrho$ we therefore need to solve
\begin{equation}
 \mathcal{S}(\varrho):= \min \limits_{\lbrace p_i, \ket{\psi_i}\rbrace}  \max \limits_{i} \mathcal{S}(\ket{ \psi_i}),
\end{equation}
where $p_i$ and $\ket{\psi_i}$ give valid decompositions of $\varrho$. Note that if we insert a pure state $\varrho= \ket{\psi}\bra{\psi}$ in the above definition we end up with the Schmidt rank $\mathcal{S}(\ket{\psi})$ of that state. 

\begin{figure}[t]
	\centering
	\includegraphics[width=0.95\columnwidth]{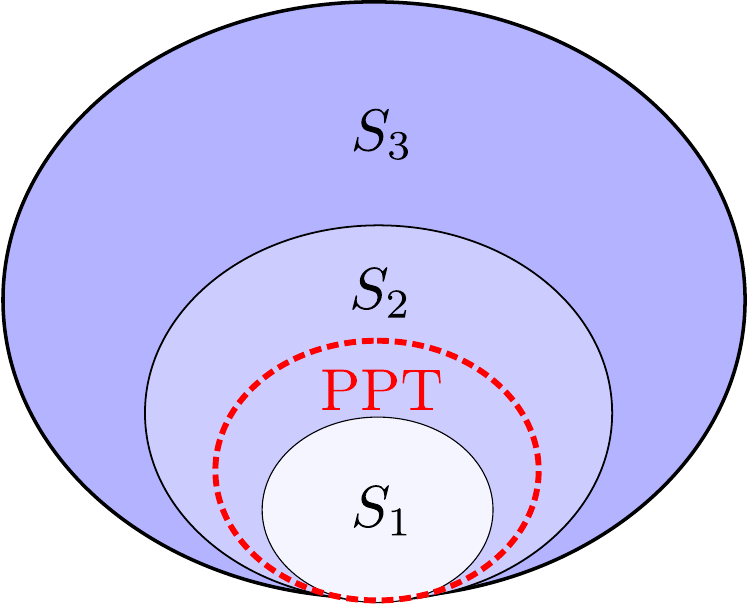}
	\caption{
	Schematic view of the state space of $3\times 3$ systems:
	The sets of states with Schmidt number less or equal $K$ are convex. 
	 The conjecture from Ref. \cite{schmidt} that for dimension $d=3$ there are no PPT states with Schmidt number $3$ was proven in Ref. \cite{YANG2016}. $S_1$ is the set of separable states. }
	\label{fig:IntroSchmidtNumber}
\end{figure}

We denote the states that have Schmidt numbers equal to or less than $K$ by $S_K$.
The set of separable states, for example, can be denoted by $S_1$.
In Fig. \ref{fig:IntroSchmidtNumber} there is a schematic overview over these sets and the set of states with positive partial transpose, which we will now cover. Note that we depicted the PPT states inside the states with Schmidt number $2$. This is true for dimension $d=3$ as it was conjectured in Ref. \cite{schmidt} and proven in Ref. \cite{YANG2016}.  \\\\

\paragraph{PPT criterion and bound entanglement:}  The PPT criterion to determine if a quantum state is separable was introduced in Ref. \cite{PhysRevLett.77.1413}. 
In Ref. \cite{Horodecki1996} it was even shown that for $(2\times2)$- and $(2\times3)$- dimensional systems the criterion is sufficient and completely solves the question whether a given state of that dimension is entangled or not. In higher dimensions states can be PPT and entangled \cite{HORODECKI1997333}.
 The PPT criterion states that for a separable state $\varrho=\sum_{i,j,k,l} \varrho_{i,j,k,l} \ket{ij } \bra{ k l}$ the partial transposition
$$
\varrho^{T_B} = \sum_{i,j,k,l} \varrho_{i,l,k,j} \ket{ij } \bra{ k l}.
$$ is positive semi-definite.

PPT entangled states are particularly interesting, since they are bound entangled, i.e., one cannot distill maximally entangled, pure singlet states via local operations and classical communication (LOCC) from them \cite{PhysRevLett.80.5239}.  Bound entangled states were first described in Ref. \cite{PhysRevLett.80.5239} and even today there are a multitude of open problems about their properties. It is, for example, not known if there are bound entangled states with negative partial transpose~\cite{OpenPr,PRXQuantum.3.010101}. Furthermore, it has been researched how these states can be used in typical quantum information theory tasks. Although they cannot be directly used for quantum teleportation, it was shown in Ref. \cite{PhysRevLett.82.1056} that they can be \textit{activated} and then perform these tasks in conjunction with other states. 
\\\\

\paragraph{CCNR criterion:} Another important separability criterion is the computable cross norm or realignment (CCNR) criterion \cite{CCNR2,CCNR}.
For the set of linear operators in the local Hilbert space $\mathcal{H}_A$ there exist local orthogonal bases $G_k^A$. The mentioned orthogonality is with respect to the Hilbert-Schmidt scalar product, i.e.,
\begin{equation}
 \Tr(G_k ^A G_j ^A )= \delta _{k,j} \quad \forall k,j \in \lbrace1,\ldots, d^2\rbrace,
\end{equation}
where $d=\text{dim}(\mathcal{H}_A)$.
An example of a local orthogonal basis would be the generators of ${\rm SU}(d)$ appropriately normalized, together with the normalized identity matrix.

Now, by performing a singular value decomposition in operator space, any density matrix $\varrho$ can be written as 
\begin{equation}\varrho = \sum_k \lambda_k G_k^A\otimes G_k^{B}.
\end{equation}
The CCNR criterion states that if $\varrho$ is separable, then $\sum_k \lambda_k\leq 1$.

In the formulation from Ref.~\cite{CCNR} and Ref. \cite{CCNRcomment} the criterion states that separable states $\varrho$ obey  
\begin{equation}
\Tr \big[\sqrt{ R(\varrho)^{\dagger} R(\varrho)}\big]\leq 1,
\end{equation}
 where $R(\varrho)$ denotes the realigned matrix given by $R(\varrho)= \sum \limits_ {i,j,k,l} \varrho_{ij,kl} \ket{ik}\bra{jl}$. 
Using this criterion it is often possible to prove that a state is entangled although it is PPT. 
\\\\

\paragraph{Entanglement quantification:} A frequent question in quantum information theory is how strong the entanglement between particles in a certain state is \cite{entanglement}. Such entanglement quantification is important, because one often thinks of entanglement as a kind of useful resource for certain tasks in quantum information processing.

Here we concentrate on the entanglement measure named linear entropy $E_{\text{lin}}$ \cite{Rungta2001}. For pure states it is defined as
\begin{equation}
E_{\text{lin}}(\ket{\psi})= 2[1- \Tr( \varrho_A^2)], \label{Elin}
\end{equation}
where the second term $\Tr( \varrho_A^2)$ is the purity of the reduced state $\varrho_A=\Tr_{B}( \ket{\psi}\bra{\psi})$. We extend this definition to mixed states with the convex-roof extension

\begin{align}
E_{\text{lin}}(\varrho)= & \inf_{\lbrace p_k, \ket {\psi_k} \rbrace}  \sum_k p_k E_{\text{lin}}(\ket{\psi_k}\bra{\psi_k}), 
\end{align}
where the infimum is calculated over all decompositions of $\varrho$. A closely related entanglement measure is the concurrence, where one considers the square root of the expression in Eq.~\eqref{Elin}.
For two qubits there is an analytical formula for the concurrence \cite{PhysRevLett.80.2245}. For larger Hilbert spaces the computation of the convex-roof expansion is often a difficult task. In Ref. \cite{toth2015evaluating} a method based on semidefinite programming was developed to compute these entanglement measures. This was utilized in Ref. \cite{Main} to calculate the linear entropy for a family of bound entangled qutrit states.

For a pure state $\ket{\psi} \in \mathcal{H}_A \otimes \mathcal{H}_B$ with $\psi_{jk} := \BraKet{j\: k}{\psi}$ the formula
\begin{align}
E_{\text{lin}}=  \sum _{jklm} \abs{\psi_{jk} \psi_{lm} - \psi_{jm }\psi_{lk} }^2\label{ElinTrick}
\end{align}
was derived for the linear entropy in Refs. \cite{Albeverio_2001,Horodecki2002}.
\\\\

\paragraph{Symmetries:} Consider maps of the form 
\be
\varrho \mapsto \sum_k p_k \big( U_k \otimes V_k \big) \varrho \big( U_k^\dagger \otimes V_k^\dagger\big), \label{localUnitarySum}
\ee
where $U_k$ and $V_k$ are unitaries acting on the respective local Hilbert spaces. This operator defines an LOCC protocol, since the unitaries $U_k$ and $V_k$ act locally and with classical communication one can arrange to apply, for example, $ U_j \otimes V_j$ with some probability $p_j$.
Since this map corresponds to an LOCC protocol, we know that it only can decrease entanglement.
The families of symmetric states we are interested in are states that are invariant under some of the above maps. Then if one has some insight into the entanglement properties for the family of symmetric states, one can obtain a lower bound for the entanglement of a general state. This is because the twirling map 
\begin{equation}
\mathcal{T}(\varrho)= \int_{ \mathcal{G}} \text{d} g  \text{ }  g \varrho  g^\dagger 
\end{equation}
maps general states to states which are invariant under the group of symmetries $\mathcal{G}\subseteq \lbrace U\otimes V | U,V \in \mathcal{U}(\mathcal{H}) \rbrace $. Twirling can be viewed as averaging over the symmetries $\mathcal{G}$ and behaves like the maps in Eq.~$\eqref{localUnitarySum}$.
This discussion shows that to get insight into the entanglement of the whole state space it is a good idea to study various families of symmetric states, where each family is determined by a group of symmetries $\mathcal{G}$. Multiple families of symmetric states have already been researched. The Werner states \cite{Werner} and the isotropic states \cite{IsotropicStates} are well known examples of such families. Other families that fulfill this property and have been extensively studied include the graph-diagonal states \cite{graphdiagonal,graphdiagonal2,graphdiagonal3} and the Greenberger-Horne-Zeilinger diagonal states \cite{GHZDiag,GHZDiag2,GHZDiag3,GHZDiag4,GeomMeasMulti}.

\section{Introduction of the Explored Family of States}
\label{sec:Family}

Now we consider states with the following symmetries:
\begin{itemize}
\item[(a)] Simultaneous cyclic permutations of the basis elements of both parties, i.e., $ \ket{ij}\bra{kl} \mapsto \ket{i\oplus n , j \oplus n}\bra{k \oplus n, l\oplus n} $ with $n \in \{1,\ldots,d-1 \}$. For dimension $d=3$, exemplary matrices $U_k,V_k$ from Eq.~\eqref{localUnitarySum} that correspond to this symmetry are
\begin{equation}
U_1\otimes V_1= \begin{pmatrix}
0&0&1\\
1&0&0\\
0&1&0
\end{pmatrix}
\otimes \begin{pmatrix}
0&0&1\\
1&0&0\\
0&1&0
\end{pmatrix}.
\end{equation}

\item[(b)]Simultaneous local phase rotations of the form $S(\varphi_1, \varphi_2,\ldots, \varphi_{d-1})= e^{i \sum_j \varphi_j \mathfrak{g}_j } \otimes e^{-i \sum _j \varphi_j \mathfrak{g}_j }$, where $\varphi_1,\ldots,\varphi_{d-1}$ are real parameters and $\mathfrak{g}_j$ are the generators of SU($d$) that are diagonal matrices. For $d=3$ this yields
\begin{align}
&U(\phi_1, \phi_2)\otimes V(\phi_1, \phi_2) = \begin{pmatrix}
e^{2\pi i \phi_1}&0&0\\
0&e^{2\pi i \phi_2}&0\\
0&0&e^{-2\pi i( \phi_1+\phi_2)}
\end{pmatrix}\\&
\otimes \begin{pmatrix}
e^{-2\pi i \phi_1}&0&0\\
0&e^{-2\pi i \phi_2}&0\\
0&0&e^{2\pi i (\phi_1+\phi_2)}
\end{pmatrix},
\end{align}
with real phases $\phi_1,\phi_2\in \mathbb{R}$. The last diagonal entry is chosen in a way that the determinant is one. This can be done, since global phases have no physical meaning. 
\end{itemize}
These states were introduced in Ref. \cite{Main} and mainly studied for dimension $d=3$. 
For a subfamiliy Sentís et al. \cite{Main} explicitly calculated the convex roof extension of the linear entropy as well as the concurrence. This is particularly interesting, because the family contains a sizable part of bound entangled states. The family is obtained by relaxing the symmetries from the axisymmetric states in Ref. \cite{Axisymmetric}. The $3\times3$ family of bound entangled states from Horodecki et al. \cite{PhysRevLett.82.1056} are also contained in the family.

To address the components of the density matrices $\varrho^{\diamond}$  that are invariant under these symmetries we use the following variables
\begin{align}
\varrho^{\diamond}_{kj, kj}=: x_{(j-k \text{ mod } d)+1},
\end{align}
and for $k\neq j$,
\begin{align}
\varrho^{\diamond}_{kk,jj}=: y_{(k-j \text{ mod } d)}.
\end{align}
For $d=3$ we have the matrix 

\NiceMatrixOptions{code-for-first-row = \scriptstyle,code-for-first-col = \scriptstyle }
\setcounter{MaxMatrixCols}{12}
\begin{equation}\varrho^{\diamond}=\begin{pNiceMatrix}[last-row,last-col,xdots/line-style={dashed}]
x_1& & & \Vdots & &y_2 & & \Vdots &&&y_1 \\
&x_2 & &  & & & &  \\
& & x_3&  & & & & \\
& & &  & & & &  & & &  &  \\
& & &  &x_3 & & &  & & &  &  \\[-6pt]
y_1& & &\Vdots & & x_1 & & \Vdots &&&y_2\\
& & &  & & & x_2&  & & &  &  \\
 & & &  &  & &  &  & & &  &  \\
& & &  & & & &  &x_2 & &  &  \\
& & & & & & & & & x_3 &  \\[-6pt]
y_2& & & \Vdots & &y_1 & & \Vdots & & & x_1 \\
 & & &  & & & &  \\
\CodeAfter
\tikz \draw [dashed] (4.5-|1) -- (4.5-|last) ;
\tikz \draw [dashed] (8.5-|1) -- (8.5-|last) ;
\end{pNiceMatrix}.
\end{equation}

In the Appendix, Sec. \ref{sec:FullMatrix}, there is a depiction of the matrix for arbitrary dimension.

The matrix has to be Hermitian in order to still be a density matrix. Therefore the parameters $y_{i}$ need to fulfill $y_{i}= y_{d-i}^*$. From the positive semi-definiteness it follows directly that $x_i\geq 0$.

 We want to find further conditions for the off-diagonal parameters, so that $\varrho^\diamond$ is positive semi-definite. For this we change the order of the basis vectors, so that the resulting matrix has a circulant block in the upper left corner and the rest of the matrix is diagonal. For $d=3$ we obtain
\NiceMatrixOptions{code-for-first-row = \scriptstyle,code-for-first-col = \scriptstyle }
\setcounter{MaxMatrixCols}{12}
\begin{equation}\varrho^{\diamond}=\begin{pNiceMatrix}[last-row,last-col,xdots/line-style={dashed}]
x_1& y_2&y_1 & \Vdots & & & & \Vdots &&& \\
y_1&x_1 &y_2 &  & & & &  \\
y_2& y_1& x_1&  & & & & \\
& & &  & & & &  & & &  &  \\
& & &  &x_2 & & &  & & &  &  \\[-6pt]
& & &\Vdots & & x_2 & & \Vdots &&&\\
& & &  & & & x_2&  & & &  &  \\
 & & &  &  & &  &  & & &  &  \\
& & &  & & & &  &x_3 & &  &  \\
& & & & & & & & & x_3 \\[-6pt]
& & & \Vdots & & & & \Vdots & & & x_3 \\
 & & &  & & & &  \\
\CodeAfter
\tikz \draw [dashed] (4.5-|1) -- (4.5-|last) ;
\tikz \draw [dashed] (8.5-|1) -- (8.5-|last) ;
\end{pNiceMatrix}.
\end{equation}

 In the upper left corner there is a circulant matrix. Its eigenvalues are given by~\cite{horn2012matrix}:
 $\lambda_{j}= x_{1} + \sum_{k=1}^{d-1}  y_{k} 
\omega^{jk} y_{k} 
 $, where 
$\omega=\rme^{\frac{2\pi \rm i}{d}}$ is the $d$-th root of unity.
 Therefore, since $\lambda_j \geq 0$ the parameters for the off-diagonal elements need to fulfill the inequality
\begin{align}
\lambda_j= x_{1} + \sum_{k=1}^{d-1} e^{\frac{2 \pi i j k}{d}} y_{k} \geq 0 
\quad \forall j\in \lbrace 0,\dots , d-1 \rbrace .
 \end{align}

In the following, we show that the physical states obeying the symmetries form a polytope in the sense that every state {that is contained in it}
can be written as convex combination   of a finite set of extremal states. These extremal states cannot be written as a non-trivial convex combination of other states in the family.  
We begin by searching
for these extremal points or vertices. 

First we note that states where for a $k\in \lbrace 2,\ldots,d\rbrace$ the parameter $x_k = \frac{1}{d}$ and all the other parameters are zero are extremal.
To find all vertices with $x_1\neq 0 $ we have a closer look at the eigenvalues of the circulant block.

With the convention $y_{0}:= x_{1}$ we have
\begin{equation}
  \sum_{k=0}^{d-1} e^{\frac{2 \pi i j k}{d}} y_{k} = \lambda_j \geq 0 \quad  \text{and } \quad  \sum_{j=0}^{d-1} \lambda_j= dy_0 .
\end{equation}
Therefore we have a polytope in the space of the possible eigenvalues.   
Now we can apply the reverse discrete Fourier transform
\begin{equation}
y_k =\frac{1}{d}\sum_{j=0}^{d-1} \lambda_j e^{\frac{-2\pi i jk }{d}}
\end{equation}
 to obtain the off-diagonal parameters of our family again. Since the reverse discrete Fourier transform is just a linear operation we conclude that we have a polytope with a finite set of vertices. For these new vertices we have $\lambda_j= dy_0=1$ for only one $j\in \{0,\ldots,d-1\}$ and the other $\lambda_k$ are zero.
  In total we have $d+ d-1= 2d-1$ vertices. Because we have to enforce the normalization condition every state is determined by $2(d-1)$ real parameters. This is due to the fact that all introduced vertices are already Hermitian.

One particular vertex corresponds to the maximally entangled state $ \ket{\phi^+_d}= \frac{1}{\sqrt{d}} \sum_i \ket{ii} $. There, $\lambda_0$ is maximal and the other eigenvalues are zero. The values of the off-diagonal parameters $y_j$ are equal to $x_1=\frac{1}{d}$ and the other $x_k$ are zero. In later sections we will restrict ourselves to the facet, where $\ket{\phi^+_d}\bra{\phi^+_d}$ is the only non-diagonal vertex.
\\\\

 \paragraph{PPT and CCNR criteria for the family.} To investigate the entanglement properties of the family $\varrho^{\diamond}$ we apply the CCNR and the PPT criterion. The partial transpose of a matrix like $\varrho^{\diamond}$ is block diagonal with $2\times2$ blocks after changing the order of the basis.
The PPT criterion states that the determinant of these blocks is, for all $j$ and $k$, positive for separable states. So, PPT states need to fulfill:
\begin{align}&      x_{j-k+1}\times x_{k-j +d+1}-\abs{y_{j-k}}^2 \geq 0 \\
&\forall j,k\in\lbrace 1,\ldots,d\rbrace, \quad j>k\nonumber
\\ \Leftrightarrow  &     \sqrt{ x_{j-k+1}\times x_{k-j +d+1}} \geq \abs{ y_{j-k} }  \\
& \forall j,k\in\lbrace 1,\ldots,d\rbrace, \quad j>k \nonumber \\
\Leftrightarrow &    \sqrt{ x_{i+1}\times x_{ d+1-i}} \geq \abs{ y_{i} } \quad  \forall i \in \lbrace 1,\ldots,d-1 \rbrace .  
\label{PPTZn} 
\end{align}

To search for PPT entangled states, we also calculate the CCNR criterion to detect entanglement.
The realigned matrix has non zero entries at the same places as the original density matrix $\varrho^{\diamond}$.
To calculate its eigenvalues we consider again a matrix where the basis vectors are ordered such that in the upper left corner there is a circulant matrix,
for which we have formulas for the eigenvalues. The sum of the absolute value of all the eigenvalues is 
\begin{equation}
\sum\limits_i  \abs{\eta _ i} = d\sum\limits_ {j} \abs{y_j} + \sum\limits_ {j= 0 }^{d-1}\bigg|\sum _{ k=0}^{d-1 } x_{k+ 1} e^{\frac{2\pi i kj}{d}} \bigg|  \label{CCNRZd}
\end{equation} 
and the CCNR criterion states that for separable states $\sum\limits_i  \abs{\eta _ i}\leq 1$.
In order to obtain explicit results from Eqs. \eqref{PPTZn} and  \eqref{CCNRZd} we only consider certain facets of the polytope.

\section{Bloch representation of the states in the family}

The goal of this section is twofold. On the one hand, we elucidate
the relation between the magic simplex~\cite{Baumgartner2006,Bertlmann2008a} 
and the family of  axisymmetric states as well as the facet of the relaxed axisymmetric states that we will study in detail. On the other hand,
we show that there exists a straightforward operator decomposition for 
these states, which is important for entanglement detection by means of the CCNR criterion.

\subsection{Axisymmetric states as subset of the magic simplex}
The main ingredient in order to discuss the magic simplex is 
a basis of maximally entangled states in the $(d\times d)$-dimensional
Hilbert space. Such bases can always be constructed~\cite{Werner2000}
(discussed below).  The magic simplex then is the set of all convex combinations
of these basis states. It is characterized by 
$d^2-1$ real parameters, 
in contrast to the full state space with $d^4-1$ parameters.

A basis of maximally entangled states can be found 
by starting from the standard Bell state in $d\times d$ dimensions,
\begin{align}
        \ket{\phi^+_d}\ =\ \frac{1}{\sqrt{d}}\sum_{j=0}^{d-1} \ket{jj} .
\label{eq:Phi+}
\end{align}
By applying local unitaries to $\ket{\phi^+_d}$ we can achieve a complete
basis. To this end, we introduce the unitary Weyl matrices $Z$ and 
$X$ (see, e.g., Refs.~\cite{Weyl1928,Vourdas2004}) that act on 
$d$-dimensional Hilbert spaces and have the
properties $Z\ket{j}=\omega^j\ket{j}$ and $X\ket{j}=\ket{j\oplus 1}$. As mentioned before,
$\omega=\rme^{\frac{2\pi\rmi}{d}}$ and the addition has to be understood
modulo $d$. Then, a full orthonormal basis $\{\ket{\phi_{kl}}\}$
of maximally entangled states is obtained by defining
\begin{align}
      \ket{\phi_{kl}}\ =\ Z^k\otimes X^l\ \ket{\phi^+_d}\ =\
                        \frac{1}{\sqrt{d}}\sum_{j=0}^{d-1}
                   \omega^{jk}\ket{j\ (j\oplus l)} ,
\label{eq:Bellbasis}
\end{align}
where $0\leq k,l \leq d-1$. For $k=l=0$ we have 
$\ket{\phi_{00}}\equiv\ket{\phi^+_d}$.

The magic simplex, i.e., the convex combinations 
$\sum_{kl} c_{kl} \ket{\phi_{kl}}\bra{\phi_{kl}}$ with
$0\leq c_{kl}\leq 1$ and $\sum_{kl}c_{kl}=1$ is still a rather complicated object.
Therefore one
may consider peculiar families of highly symmetric states that are subsets
of the magic simplex, but are described by fewer parameters. The first
family we consider here are the axisymmetric states~\cite{Eltschka2015}
that generalize the isotropic states~\cite{IsotropicStates}, 
\begin{align}
   \varrho_{\mathrm{axi}}\ =\ p \ket{\phi^+_d}\bra{\phi_d^+} +
                           q\ \varrho_0 + (1-p-q)\ \varrho_{(1)}
                            ,
\label{eq:axi}
\end{align}
with $ 0\leq p, q\leq 1$ where
\begin{align}
     \varrho_0\ =\ &  \frac{1}{d-1}\ \sum_{k=1}^{d-1}
                    \ket{\phi_{k0}}\bra{\phi_{k0}}
                    \nonumber
\\
            =\ & \frac{1}{d-1} \sum_{k=1}^{d-1} \ket{kk}\bra{kk}\ -\
                   \frac{1}{d(d-1)} \sum_{k\neq l} \ket{kk}\bra{ll} ,
\label{eq:axi-upper-left}               
\end{align}
and
\begin{align}
     \varrho_{(1)}\ =\ &  \frac{1}{d(d-1)}\ \sum_{k=0,l=1}^{d-1}
                    \ket{\phi_{kl}}\bra{\phi_{kl}}
                    \nonumber
\\
            =\ & \frac{1}{d-1} \sum_{l=1}^{d-1} 
    \frac{1}{d}\sum_{k=0}^{d-1}\ket{k\ (k\oplus l)}\bra{k\ (k\oplus l)}
 .
\label{eq:axi-lower}               
\end{align}
The axisymmetric states are all the mixed states that share the symmetries
of the Bell state $\ket{\phi^+_d}$ (simultaneous local phase rotations 
with opposite
sign for the two parties, simultaneously exchanging  the labels of the levels
of the local $d$-state systems). They are convex combinations of 
the three states given above for all local dimensions $d$.
Evidently this family is a subset of the magic simplex.
It was shown~\cite{Eltschka2015} that there are no
bound entangled axisymmetric states, so in order to have such states in the family we have to lower the degree of symmetry.

It turns out that it is sufficient to relax the symmetry under arbitrary simultaneous level permutations of the two parties into a cyclic permutation symmetry in order to allow for bound entanglement. One consequence of this modification compared to the axisymmetric states is that the state $\varrho_{(1)}$ in Eq.~\eqref{eq:axi-lower} splits up into $d-1$ states $\varrho_l$, $l=1\ldots d-1$. 
Another consequence of the symmetry relaxation is that the offdiagonal elements may have different modulus as well as complex phases, as we explained in the previous section. We restrict our attention to a subfamily of the relaxed axisymmetric states that evidently lies within the magic simplex.  The peculiar facet 
that we consider here is the convex hull of the standard Bell state, 
$\ket{\phi^+_d}$, and the states $\varrho_l$ that result from splitting up $\varrho_{(1)}$,
\begin{equation}
   \varrho^\diamond _{\mathrm{facet}}\ =\ p\ \ket{\phi^+_d}\bra{\phi_d^+}\ +\
      \sum_{l=1}^{d-1}  q_l\ \varrho_l,
      \label{eq-rhofacet}
\end{equation}
where $p+\sum_l q_l = 1$, and
\begin{align}\label{eq:rhol}
      \varrho_l\ =\    &  \frac{1}{d}\ \sum_{k=0}^{d-1}
                    \ket{\phi_{kl}}\bra{\phi_{kl}}
                    \nonumber
\\
            =\ &\frac{1}{d}\sum_{k=0}^{d-1}
         \ket{k \ (k\oplus l)}\bra{k\  (k\oplus l)}
          , \ \ \forall l\in \{1\ldots d-1\}
 .
\end{align}

\subsection{Bloch representation for a facet of the relaxed axisymmetric states}
Because of their high symmetry it is per se interesting to study what this means for the Bloch representation of the states in the family. It will turn out particularly useful for the application of the CCNR criterion.

As a matrix basis for the Bloch representation we will choose the (unitary) displacement operators $D_{jk}$ with their standard definition~\cite{Vourdas2004},
\begin{align}
    D_{jk}\ =\ Z^j\ X^k\ \omega^{-\frac{jk}{2}}
     .
\label{eq:displacement}
\end{align}
An important rule for calculations is
$X^k Z^j=Z^jX^k\omega^{-jk}$.
Again we start with the standard Bell state whose Bloch representation is given by
\begin{align}
    \ket{\phi^+_d}\bra{\phi^+_d}\ =\ 
    \frac{1}{d^2} \sum_{a,b=0}^{d-1}
                       D_{ab}\otimes D_{ab}^*
                        .
\label{eq:Bell-Bloch}
\end{align}
The validity of this equation is readily seen from
the general representation of the 
$\mathsf{SWAP}$ operator valid for any orthonormal
matrix basis $\{G_j\},\ j=1\ldots d^2, 
\Tr\big(G_j^{\dagger}G_k\big) =d\delta_{jk}$, that is,
$\mathsf{SWAP}=\frac{1}{d}\sum_{j}G_j\otimes
                         G_j^{\dagger}$, and the
well-known relation $\mathsf{SWAP}=d\ket{\phi^+_d}\bra{\phi^+_d}^{T_B}$
(cf.~Ref.~\cite{WolfLectureNotes}).
With the definition \eqref{eq:Bellbasis} we obtain also the Bloch representations of the other states of the basis $\{\ket{\phi_{kl}}\}$,
\begin{align}
    \ket{\phi_{kl}}\bra{\phi_{kl}}\ =\ 
    \frac{1}{d^2} \sum_{a,b=0}^{d-1}
                       D_{ab}\otimes D_{ab}^*\ \omega^{al+bk}
                        .
\label{eq:Bellbasis-Bloch}   
\end{align}
It is now straightforward to determine the Bloch decomposition of the remaining extremal states of the facet, Eq.~\eqref{eq-rhofacet}. We find
\begin{align}
 \varrho_l \ =\ &  \frac{1}{d^2}\  \sum_{a=0}^{d-1}\           D_{a0}\otimes D_{a0}^*\ \omega^{al}
                        \ ,  \ 
\forall   l\in \{1\ldots d-1\}
                        .
\end{align}
These results are remarkable because of their simplicity. The most important property is that the Bloch decomposition of the axisymmetric states is diagonal in the Weyl basis, which facilitates direct application of the CCNR criterion: It suffices to use the absolute values of the Bloch coefficients instead of the singular values of the operator Schmidt decomposition (see Sec.~II).

\section{Entanglement analysis for a certain facet}
\label{SectionWithFacet}

\begin{figure}[t]
	\centering
	\includegraphics[scale=0.27,trim={4cm 8cm 3cm 3cm},clip]{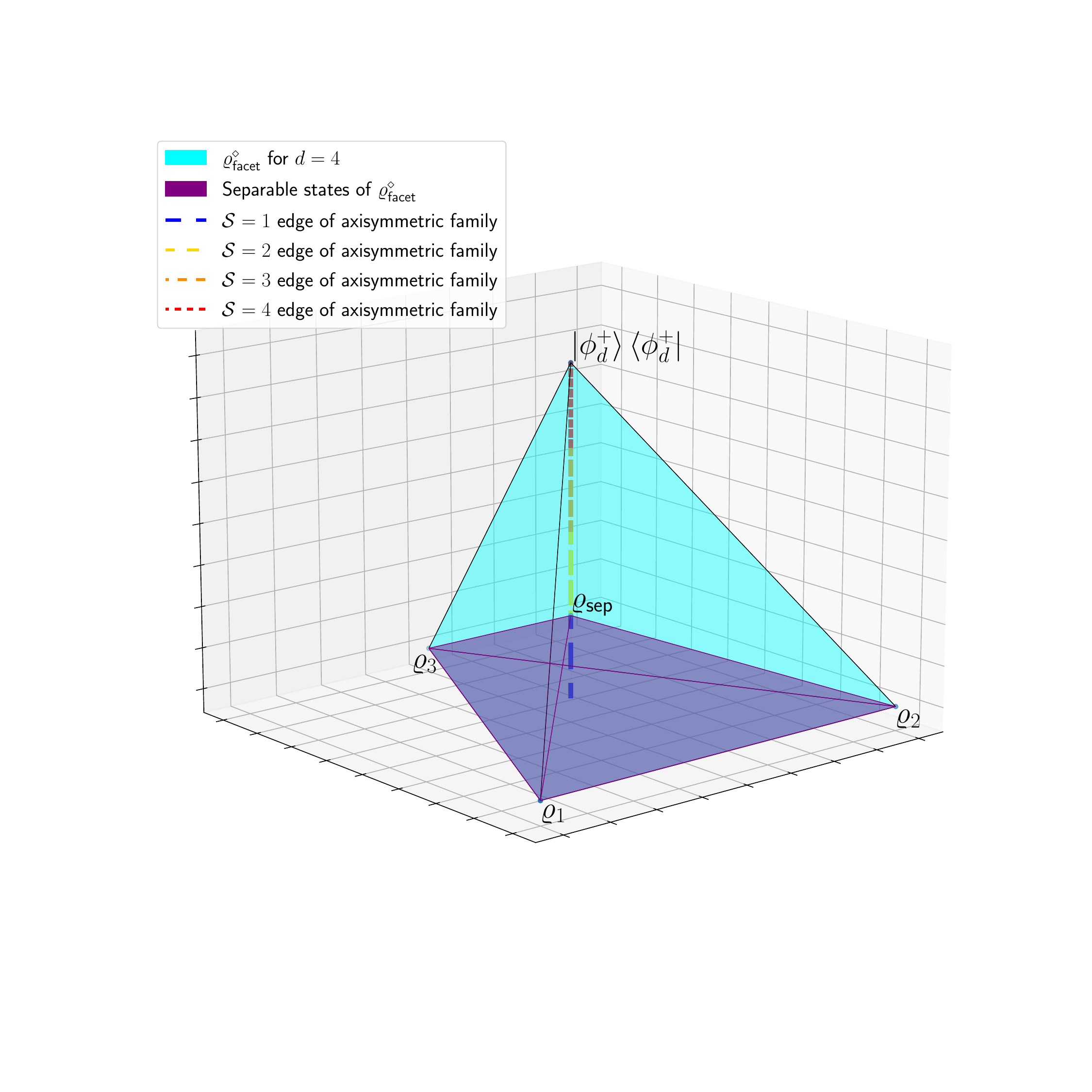}
	\caption{Facet for $d=4$. If $x_2= x_3=x_4$ we arrive at an edge from the axisymmetric states from Ref. \cite{Axisymmetric}. We will prove that all separable states lie in the purple (dark gray) polytope. }
	\label{img:grafik-dummy}
\end{figure}

We will fully characterize the set of separable states in a subfamily of $\varrho^{\diamond}$ corresponding to a facet of the polytope. By adapting the state parameters to those of Sec.~\ref{sec:Family}, we obtain for the states in the facet of the polytope
\begin{equation}
\varrho_{\text{facet}}^\diamond:=d x_1\ket{\phi^+_d}\bra{\phi^+_d} + \sum_{k=1}^{d-1} x_{k+1} \sum_ {j=0}^{d-1}  \ket{j\;( j\oplus k)}\bra{ j\;( j\oplus k)}\,, \label{ImportantFacet}
\end{equation}
which are determined by $d-1$ real parameters. A schematic overview of the facet can be seen in Fig. \ref{img:grafik-dummy}. How one obtains results for the Schmidt numbers will be discussed in Sec. \ref{sec:SCHMIDT}. 

The previously defined off-diagonal parameters $y_k$ are all set to $y_k= x_1$.
 The normalization relation is still the same:
\begin{equation}
d \sum_{k=1}^d x_k = 1.
\end{equation}
We want to show that in this facet all separable states are convex combinations of $\varrho_{k}:= \frac{1}{d}\sum_ {j=0}^{d-1}  \ket{j\;( j\oplus k)}\bra{ j\;( j\oplus k)}$ and the state $\varrho_{\text{sep}}$, which is defined such that $\frac{1}{d^2}= x_{j}= y_{j} \forall j \in \lbrace 1,\ldots,d \rbrace$.
The state $\varrho_{\text{sep}}$ is part of the family of axisymmetric states in Ref.~\cite{Axisymmetric}. There it was shown to be separable. Another proof is given in Ref.~\cite{vidal1999robustness} 
[see Eq.~(B5) therein]. 
Additionally we show a decomposition into separable states of $\varrho_{\text{sep}}$ in the Appendix, Sec. \ref{Sec:Decomp}.
\\\\

 \paragraph{Linear entropy as entanglement criterion.} In Ref.~\cite{Main} a method was described to calculate the linear entropy of states that are invariant under an entanglement preserving symmetry. It can be applied as follows. The set $\lbrace \ket{ \phi_d^+}\bra{\phi_d^+} , \varrho_{1}, \ldots, \varrho_{d-1} \rbrace$ consists of states that span the facet we are interested in. They are invariant under the symmetries, which were used to construct these states. We can parametrize an arbitrary state as follows:
\begin{align}\nonumber
  & \Tr( \varrho \varrho_k)=  x_{k+1} \quad \forall k \in \lbrace 1, \ldots,d-1 \rbrace  \\  \quad \wedge \quad & \Tr( \varrho \ket{\phi_d^+}\bra{\phi_d^+} )= d x_1  . 
\end{align}
A state in the facet $ \sigma_{x_1 ,\ldots, x_d}$  is uniquely determined by the parameters $x_1,\ldots, x_d$. With the theorem from Ref.~\cite{Main} we can get a formula for the linear entropy. Explicitly we have to do the following steps:

\begin{enumerate}
\item Find a parametrization for all pure states in the span of $\sigma$ in dependence of the facet parameters $x_1 ,\ldots, x_d$ and other parameters $\xi$. Denote these pure states by $ \ket{ \psi_{\sigma}}$. It follows that every decomposition of a state $\sigma$ only uses pure states from $ \ket{ \psi_{\sigma}}$. 
\item Compute the function $ \tilde{E}_{\text{lin}}( x_1, \ldots, x_d)= \min_{\xi} E_{\text{lin}}( \ket{\psi_{\sigma}(x_1,\ldots,x_d, \xi)})$, where $\xi$ are the parameters of $\ket{ \psi_{\sigma}}$ that are not facet parameters.
\item $ \tilde{E}_{\text{lin}}( x_1, \ldots, x_d)$ is not necessarily convex. Compute its convexification $\tilde{E}^{c}_{\text{lin}}( x_1, \ldots, x_d)$. 
\item Then the identity $ \tilde{E}^{ c}_{\text{lin}}( x_1, \ldots, x_d)= E_{\text{lin}}(\sigma_{x_1 ,\ldots, x_d})$ holds.
\end{enumerate}
In the following we construct a subset $\mathcal{K}$ of the facet that includes all separable states. 
This property would be fulfilled if the following  were true
\begin{equation}
\tilde{E}^{ c}_{\text{lin}}( x_1, \ldots, x_d)=0 \Longrightarrow \sigma_{x_1 ,\ldots, x_d} \in \mathcal{K}.
\end{equation} 
We know that the zero set of the convexified function $\tilde{E}^{ c}_{\text{lin}}( x_1, \ldots, x_d) $ is just the convexification of the zero set of the initial function $\tilde{E}_{\text{lin}}( x_1, \ldots, x_d)$:
\begin{align}
&\lbrace x_1,\ldots,x_d |\tilde{E}^{ c}_{\text{lin}}( x_1, \ldots, x_d) =0\rbrace \nonumber  \\&=\lbrace x_1,\ldots,x_d|\tilde{E}_{\text{lin}}( x_1, \ldots, x_d)=0\rbrace^{c}. \label{convexArgument}
\end{align}
For now we search a set that includes all states where $\tilde{E}_{\text{lin}}( x_1, \ldots, x_d)=0$. This can be achieved by finding the states where $\tilde{E}_{\text{lin}}( x_1, \ldots, x_d)>0$, and considering all other states. \\
Pure states that build a decomposition of a state in the facet are of the form:
\begin{equation}
\ket{ \psi _{\sigma }}= \sqrt{ d x_{1}} \ket {\phi_d^+} + \sum _{k=1} ^{d-1} \sqrt{d x_{k+1}} \sum_{j=0} ^{d-1} \xi_{k,j} \ket{j\;( j\oplus k)}, \label{param}
\end{equation}
where $  \sum_{j=0}^{d-1} \abs{\xi_{k,j}}^2=1$ and $\xi_{j,k}\in \mathbb{C}$ for all $k \in \lbrace 1,\ldots, d-1   \rbrace \quad$.

Now by using Eq.~\eqref{ElinTrick} we get the following formula:
\begin{equation}
\tilde{E}_{\text{lin}}( x_1, \ldots, x_d)= \min_{\xi} \sum _{jklm} \abs{\psi_{jk} \psi_{lm} - \psi_{jm }\psi_{lk} }^2,  \label{AfterElinTrick}
\end{equation}
where $ \psi_{lm}:= \ket{ \psi (x_1,\ldots, x_d, \xi)}_{lm} $. 
The minimization over the complex phases is simple. For two complex numbers $r_1 e^{i\phi_1} $ and $r_2 e^{i\phi_2}$, the minimum of $\abs {r_1 e^{i\phi_1}+r_2 e^{i\phi_2}}^2$ is achieved for $ \phi_1= \phi_2 + \pi$. Therefore our minimization problem yields the same value if we assume all coefficients $\xi_{k,j}$ to be real. 
This allows us to conclude that, for $\tilde{E}_{\text{lin}}( x_1, \ldots, x_d)$ to be zero, there has to exist a $\xi$ so that
\begin{align}
 &  \abs { \psi_{jk}\psi_{lm}- \psi_{jm}\psi_{lk}}^2 = 0 \quad \forall j,k,l,m \in \lbrace 0, \ldots, d-1 \rbrace \nonumber \\
 \Leftrightarrow &   \psi_{jk}\psi_{lm}= \psi_{jm}\psi_{lk}\quad \forall j,k,l,m \in \lbrace 0, \ldots, d-1 \rbrace. \label{coeffequation}
\end{align}
  An implication of Eq. \eqref{coeffequation} is the following:
 \begin{equation}
  \prod _{l=1}^{ d}\psi_{ll} =\prod_{l=1}^d \psi_{l \text{  } l\oplus m}\quad \forall m \in \lbrace 0,\ldots, d-1\rbrace.
 \end{equation}
For example, for $d=3$, $ \psi _{00} \psi_{11}\psi_{22}=\psi _{01} \psi_{10}\psi_{22}=\psi _{01} \psi_{12}\psi_{20}$. 
Inserting the state coefficients from Eq. \eqref{param}, we obtain
\begin{equation}
\left(\sqrt{ x_1}\right)^d= \left( \sqrt{ d x_{k+1}}\right)^d \prod _{j=0 }^{d-1} \xi_{k,j} \leq \left(\sqrt{x_{k+1}}\right)^d.\label{ineq}
\end{equation}
Here the inequality $\prod _{j=0 }^{d-1} \xi_{k,j}\leq \left(\frac{1}{\sqrt{d}}\right)^d $ was used 
\footnote{Consider the optimisation problem: 
$$ \scriptstyle   \max_{(a_1,\ldots, a_d)\in \mathbb{R_{+}}^d}  \mathlarger{\Pi}_{i= 1 }^{d} a_i \quad\text{subject to} \quad \mathlarger{\Sigma}_i a_i^2=1.
$$
For $d=2$ the solution is straightforward. For higher dimensions one can solve this iteratively. The geometric interpretation is that the hyperrectangle with largest volume that is contained in a hypersphere is a hypersquare. 
}.

Since squaring and taking the $d$th power are both monotone operations, the inequalities \eqref{ineq} cannot be true (regardless of $\xi$) if for some $k \in \lbrace 2,\ldots,d\rbrace \quad x_1>x_k$. But since the inequalities \eqref{ineq} are an implication of $\tilde{E}_{\text{lin}}( x_1, \ldots, x_d)=0 $ we get the true statement   
\begin{equation}
\left(\tilde{E}_{\text{lin}}( x_1, \ldots, x_d)=0 \quad \Rightarrow \quad     x_1\leq x_k \quad \forall k \in \lbrace 2,\ldots,d\rbrace \right)
.\end{equation}
Therefore we know that the set of states where the coefficients have the property $  x_1 \leq x_r \quad \forall r \in \lbrace 2,\ldots,d\rbrace $ includes all the states with facet parameters so that $\tilde{E}_{\text{lin}}( x_1, \ldots, x_d)=0$. It is easy to see that the set of states with coefficients obeying $   x_1 \leq x_k \quad \forall k \in \lbrace 2,\ldots,d\rbrace $ coincides with the polytope spanned by $ \varrho_{\text{sep}}, \varrho_{1}, \ldots,\varrho_{d-1} $. According to Eq. \eqref{convexArgument} the convexification of this subset yields a set that includes all states where the linear entropy is zero. Since our subset is already convex, the convexification leaves it invariant. This shows that all separable states in the facet are convex combinations of $ \varrho_{\text{sep}}, \varrho_{1}, \ldots,\varrho_{d-1} $. But since we showed at the beginning that $\varrho_{\text{sep}}$ is separable, the converse is also true and we know that all states in the polytope spanned by  $ \varrho_{\text{sep}}, \varrho_{1}, \ldots,\varrho_{d-1} $ are separable.
\\\\

\begin{figure}[t]
	\centering
	\includegraphics[width=0.95\columnwidth,trim={0.3cm 0.cm 1.8cm 1.8cm},clip]{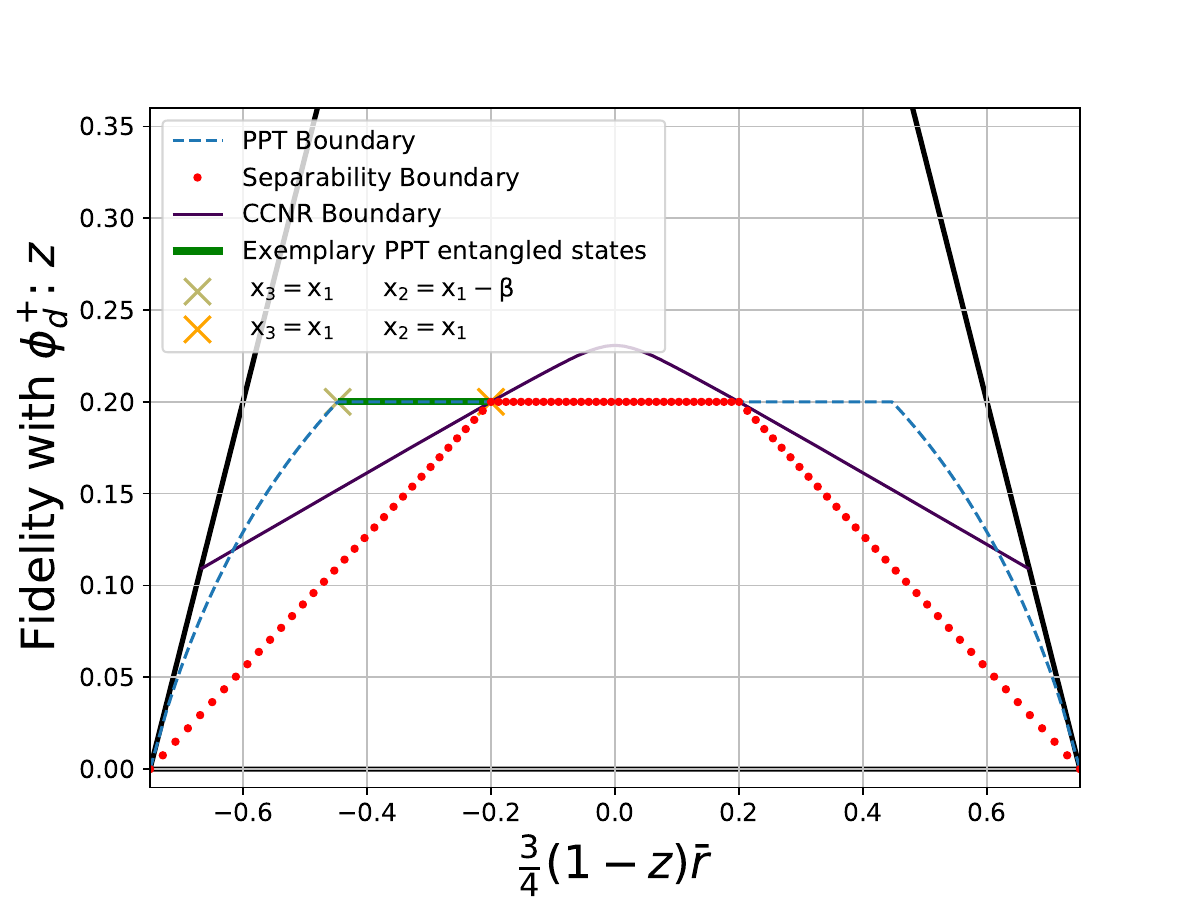}
	\caption{Cross-section of the facet in $d=4$
	in 
	$x_3= \frac{1}{4d}(1-z)$
	and 
	$x_{2,4}=\frac{3}{4d}(1-z) \frac{1\pm \overline{r}}{2} $, where $z$ and $\overline{r}$ are fidelity parameters analogous to those defined in Eqs.~\eqref{eq:Param1}-\eqref{eq:Param3}.}
	\label{img:Exemp}
\end{figure}

\paragraph{PPT and CCNR criteria for the facet.}
{From Eq. \eqref{PPTZn} we can directly state the PPT criterion for the facet:
\begin{equation}
     \sqrt{ x_{i+1}\times x_{ d+1-i}} \geq \abs{ x_{1} }\quad \forall i \in \lbrace 1,\ldots,d-1 \rbrace .  
\label{PPTZnFacet} 
\end{equation}
}
Next  we look at the CCNR criterion given in Eq. \eqref{CCNRZd} for the facet
\begin{align}
\sum\limits_i  \abs{\lambda _ i} &= d(d-1)x_1 + \sum\limits_ {j= 0 }^{d-1}\bigg|\sum _{ k=0}^{d-1 } x_{k+ 1} e^{\frac{2\pi i kj}{d}} \bigg|.
\end{align}
Applying the triangle inequality yields:
\begin{align}
\allowdisplaybreaks
\sum\limits_i  \abs{\lambda _ i}&\geq  d(d-1)x_1 + \abs{\sum\limits_ {j= 0 }^{d-1}\sum _{ k=0}^{d-1 } x_{k+ 1} e^{\frac{2\pi i kj}{d}} }  \nonumber
\\
&=  d(d-1)x_1+ \abs { dx_1+ \sum _{k=1 }^{d-1} x_{k+1} \frac{ 1- e^{\frac{2\pi i k d}{d}}}{1- e^{\frac{2\pi i k }{d}}} }  \nonumber\\
&=  d(d-1)x_1+ \abs { dx_1+ \sum _{k=1 }^{d-1} x_{k+1} \frac{ 1- 1}{1- e^{\frac{2\pi i k }{d}}} }\nonumber\\
&=  d(d-1)x_1+ \abs {  dx_1 } \nonumber \\
&=  d^2x_1. \label{eq:CCNRApprox}
\end{align}
Therefore we know that if $x_1> \frac{1}{d^2}$ the state is entangled according to the CCNR criterion. By its negation, separability implies that $ x_1 \leq \frac{1}{d^2} $, { which is in accordance with our result from the previous section. }
We now give an example for some PPT entangled states with 
\begin{align}
&x_1 \leq\frac{1}{d^2}, \quad \beta\in [0,x_1 ], \quad l \in\lbrace 2,\ldots , \lfloor  \frac{d}{2}\rfloor\rbrace \quad  \nonumber \\ &\forall k \in \lbrace 2,\ldots,d\rbrace \setminus \lbrace l, d-l +2  \rbrace \label{SomeKindOfStates} \nonumber \\ &\quad  x_k= x_1,  \quad x_l = x_1-\beta,  \quad x_{d+2-l} = \frac{1}{d} -(d-1)x_1 +\beta .
\end{align}
We calculate the CCNR criterion for these states:
\begin{widetext}
\begingroup
\allowdisplaybreaks
\begin{align}
\sum \abs{\eta_i}&= d(d-1) x_1+\frac{1}{d}+\sum_{j=1}^{d-1} \abs{ \sum _{k=0 }^{d-1} x_{k} e^{\frac{2\pi i jk}{d}}}  \nonumber\\
&= d(d-1) x_1+\frac{1}{d}+\sum_{j=1}^{d-1} \abs{  -\beta e^{\frac{2\pi i jl }{d}}+ (\frac{1}{d} -dx_1 +\beta) e^{\frac{2\pi i j(d+2-l) }{d}} }\nonumber\\
&= d(d-1) x_1+\frac{1}{d}+\sum_{j=1}^{d-1} \abs{ e^{\frac{4\pi ij (1-l) }{d}}  (\frac{1}{d} -dx_1 +\beta)  -\beta} \nonumber\\
&\geq   d(d-1) x_1+\frac{1}{d}+\sum_{j=1}^{d-1} \abs{ \abs {e^{\frac{4\pi ij (1-l) }{d}}  (\frac{1}{d} -dx_1 +\beta)}  -\beta} \nonumber\\
&=   d(d-1) x_1+\frac{1}{d}+ (d-1 ) ((\frac{1}{d} -dx_1 +\beta)  -\beta) \nonumber\\
&=   d(d-1) x_1+\frac{1}{d}+ (d-1)  (\frac{1}{d} -dx_1 )  =1,
\end{align}
\endgroup
\end{widetext}

where we applied the reverse triangle inequality. 
Since $e^{\frac{4\pi ij (1-l) }{d}}$ is not equal to 1 for all $j$, the inequality is not tight and thus we know that the state is entangled. Next we calculate for which values of $\beta$ the states are PPT. For the states described in Eq.~\eqref{SomeKindOfStates}, inequality \eqref{PPTZn} reads
\begingroup
\allowdisplaybreaks
\begin{align}
x_1 \leq& \sqrt{ (x_1 -\beta ) (\frac{1}{d} -(d-1 ) x_1 + \beta )} \nonumber \\
  &= \sqrt{ (x_1-\beta)(\frac{1}{d} - dx_1) + x_1^2- \beta^2  } \nonumber \\
  \Leftrightarrow 0\leq& (x_1-\beta)(\frac{1}{d} - dx_1) - \beta^2 \nonumber \\
   \Leftrightarrow 0\geq&   \beta^2+(\frac{1}{d} - dx_1)\beta-(\frac{1}{d} - dx_1)x_1 .
\end{align}
\endgroup
We check for which $\beta$ the equality is fulfilled:
\begin{equation}
\beta_{ \pm }=  -\frac{\frac{1}{d} - dx_1}{2}\pm \sqrt{\frac{(\frac{1}{d} - dx_1)^2}{4} +(\frac{1}{d} - dx_1)x_1  }.
\end{equation}
For the case that there is a plus sign the solution is positive ($\beta_+$) and the state is PPT entangled for $ \beta \in [0,\beta_+ ]$. The states can be seen in Fig. \ref{img:Exemp} for a cross-section in $d=4$.

In this section we restricted our analysis of entanglement properties to a facet of the state polytope. We proved that the state $\varrho_{\text{sep}}$, where all matrix elements have the same value, is separable. Moreover we proved that all separable states are in the polytope spanned by $\varrho_{\text{sep}}$ and the vertices that correspond to states with diagonal density matrices. Since the PPT criterion is easy to compute we have a full characterization of the PPT entangled states in the facet.

In the Appendix, Sec. \ref{sec:AppPPTEnt},  
we also give an example of an entangled state that is not detected by the CCNR criterion.

\section{Characterizing the dimensionality of entanglement}

\subsubsection{Calculation of Schmidt numbers}
\label{sec:SCHMIDT}
To get some insight into the dimensionality of entanglement of our states we want to calculate their Schmidt numbers.
In analogy to entanglement witnesses, Schmidt number witnesses play an important role in calculating Schmidt numbers.
In Ref.~\cite{terhal2000schmidt} Schmidt number witnesses were introduced. 
An observable $W$ is called a Schmidt witness of class $K \in\mathbb{N}$ if and only if $\Tr(W \rho)\geq 0$ for all $\rho$ with Schmidt number smaller than $K$. One also requires that at least one state with Schmidt number $K$ is detected, i.e., fulfills $\Tr(W \rho)< 0$.
A $K$-Schmidt witness $W_1$ is called finer than a $K$-Schmidt witness $W_2$  if $W_1$ detects the same states as $W_2$, and some states in addition.

It was proven in Ref.~\cite{terhal2000schmidt} that  $W= \frac{K-1 }{d} \mathbb{I} - \ket{\phi_d^+} \bra{\phi_d^+}$ is an optimal $K$-Schmidt witness. We calculate the states in the facet that are detected by this witness:
\begin{align}
\Tr(W \varrho )&= \Tr\Big[\Big(\frac{K-1 }{d} \mathbb{I} - \ket{\phi_d^+} \bra{\phi_d^+}\Big) \Big(a\ket{\phi_d^+}\bra{\phi_d^+} 
\nonumber \\
&+ \sum_{k=1}^{d-1} b_k \sum_ {j=0}^{d-1}  \ket{j\;( j\oplus k)}\bra{ j\;( j\oplus k)}\Big)\Big] 
\nonumber \\
&= \frac{K-1}{d}  - \Tr( a\ket{ \phi_d^+}\bra{\phi_d^+ }) \nonumber  \\
&= \frac{K-1}{d} -a. \label{eq:firstWitness}
\end{align} 
Since we are interested in the separable states, which have Schmidt number one, we look at the case $K= 2$.
This yields
\begin{align}
\Tr(W \varrho )&\geq 0
\quad \Leftrightarrow \frac{1}{d} -a  \geq 0  \nonumber \\
\Leftrightarrow \frac{1}{d}  &\geq dx_1  
\quad \Leftrightarrow \frac{1}{d^2}   \geq x_1,
\end{align}
which is the same bound as the one we already derived above
in Eq. \eqref{eq:CCNRApprox}. 

We will now describe a strategy to approximate the border between states with maximal Schmidt number and states with less than maximal Schmidt number. First one searches pure states that do not have full Schmidt rank and are twirled to the facet. Then the resulting 
states after the twirling operation cannot have full Schmidt number either since twirling is an LOCC map.

 Therefore we start by looking at the full Hilbert space $\mathcal{H} \otimes \mathcal{H}$ and consider how the pure states are mapped by the twirling operator. Let $\sigma \in \mathcal{H} \otimes \mathcal{H}$ and $\mathcal{T} \sigma = \varrho $. Then,
\begin{align}
& \varrho _{i\text{ } i \oplus j ,i \text{ }i \oplus j } = \frac{1}{d}\sum_k \sigma _{k\text{ } k \oplus j ,k\text{ } k \oplus j } \quad \forall i,j \in \lbrace 0,\ldots,d-1  \rbrace \\
&\varrho _{i\text{ } i  ,i \oplus j \text{ } i \oplus j } =\frac{1}{2d}\sum_k \left( \sigma _{k\text{ } k  ,k\oplus j \text{ } k \oplus j }+\sigma _{k\oplus j \text{ } k \oplus j,k\text{ } k   }\right). 
\end{align}
If the twirling operator maps the state $\sigma $ to the state $\varrho$, then we know that the Schmidt number of $\sigma$ is greater than or equal to the Schmidt number of $\varrho$, since twirling is an LOCC map. For simplicity, we investigate the problem for dimension $d= 3$. In Ref. \cite{Main} we have the following parametrization of the pure states that are mapped in the span of $\rho^\diamond$:
\begin{align}
\ket{\psi_{\sigma}}&= \sqrt{ z} \ket{\phi_d^+} + \sqrt{1-z} [ \sqrt{\frac{1+ \overline{r}}{2}}(  a\ket{01}+b \ket{12} \nonumber \\&+ c \ket{20})+\sqrt{\frac{1- \overline{r}}{2}}(  e\ket{02}+f \ket{10}+ g \ket{21}) ]. \label{paramD3}
\end{align}
Here $z$ is a facet parameter with respect to $\ket{\phi_d^+}$, $\overline{r } \in [-1,1]$, and the complex parameters $a$, $b$, $c$, 
$e$, $f$ and $g$ fulfill the normalization conditions 
$\abs{a}^2 + \abs{b}^2 + \abs{c}^2=1 $ and $\abs{e}^2 + \abs{f}^2 + \abs{g}^2=1 $. The connection to the previously used parametrization is 
\begin{align}
& x_1 =  \frac{z}{3},  \label{eq:Param1} \\
&x_2= \frac{(1-z)( 1+ \overline{r})}{6},\\ 
&x_3= \frac{(1-z)( 1- \overline{r})}{6}. \label{eq:Param3}
\end{align}
In the Appendix, Sec. \ref{sec:AppSchm}, we calculated some upper bounds for several values of $\overline{r}$ by choosing the parameters $a,\dots , g$, so that the coefficient matrix becomes circulant.

\begin{figure}[t]
	\centering
	\includegraphics[width=0.95\columnwidth]{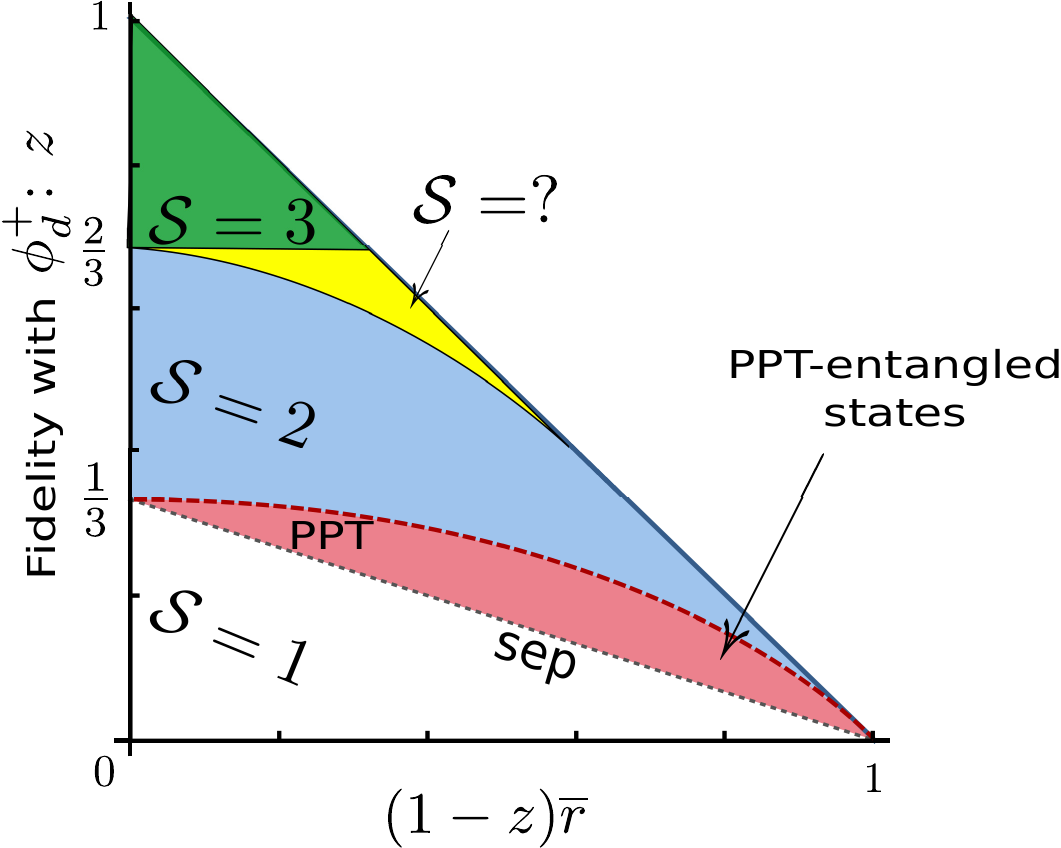}
	\caption{Progress in calculating the Schmidt numbers $\mathcal{S}$ of the states in the facet for $d=3$.
	While witnesses can give a lower bound on the Schmidt number, showing that a state with non-full Schmidt rank is twirled to a state in the facet can give an upper bound.}
	\label{img:grafik-dummy111}
\end{figure}

The obtained results for the Schmidt numbers in the facet for dimension $d=3$ are schematically summarized in Fig.~\ref{img:grafik-dummy111}. For states in the yellow area we still do not know if they have Schmidt number 2 or 3. The curve between the yellow and the blue area is given by Eq. \eqref{CurveYB} in the Appendix, Sec. \ref{sec:AppSchm}.

For higher dimensions one can also obtain upper bounds for the Schmidt number by twirling of pure states without full Schmidt rank.
We start from the parametrization of an arbitrary pure state that is twirled to the facet, which can be seen in Eq. \eqref{param}. We choose $a_{k,j} =-\frac{1}{\sqrt{d}}$, so that the coefficient matrix becomes circulant. Because of this simplification, we will not find all states with maximal Schmidt number. We again set the eigenvalue that is a sum of the distinct matrix entries to zero:
\begin{equation}
\sqrt{x_1}=\sum _{k=1 }^{d-1} \sqrt{x_{k+1}} \\
\end{equation}
In the Appendix, Sec. \ref{sec:NonFullSR}, we solve this equation for $x_1$.

With this equation we certified for some  states that they do not have full Schmidt rank. For $d=3$, we show the resulting upper bound for Schmidt number $2$ states in Fig.~\ref{fig:SDPSchmidt}. Note that applying Schmidt witnesses on the other hand gives lower bounds on the Schmidt number of the respective states.

Another idea to get insight into the Schmidt numbers is by calculating the $K$-concurrence from Ref. \cite{gour} using the convex characteristic curve method. If the $K$-concurrence then vanishes for a particular state we know that its Schmidt number is smaller than $K$. However we realized that this idea is impractical, since in the analogous formulas like Eq. \eqref{AfterElinTrick} for $K>2$ one cannot eliminate the complex phases.

\subsection{Computation of SDP hierarchy}
\begin{figure*}[t]
	\centering
	\includegraphics[scale=0.55]{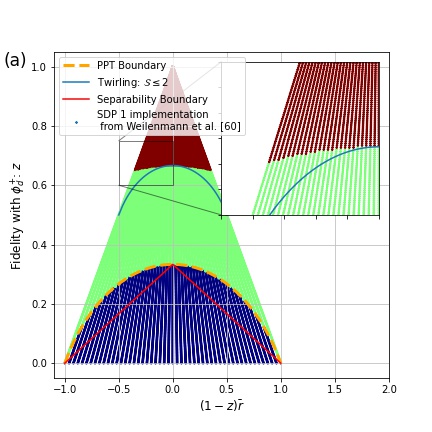}
		\includegraphics[scale=0.55]{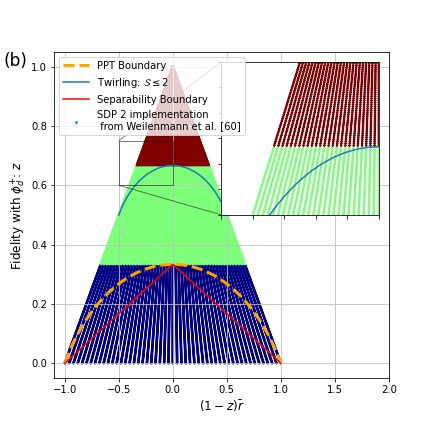}
	\caption{Calculation of the Schmidt numbers $\mathcal{S}$ of the states in the facet for $d=3$: (a) states with at least Schmidt number 1 (blue, bottom),
2 (green, middle), and 3 (red, top) and (b) the blue states in the region at the bottom are 2-unfaithful, while the green ones in the middle are
3-unfaithful. The parametrization is analogous to that in \cite{Main}. For $500\times 50$ states we calculated 
(a) SDP1 and (b) SDP2 from Ref. \cite{Weilenmann}. 
	 }
	\label{fig:SDPSchmidt}
\end{figure*}

In Ref.~\cite{Weilenmann} a method was presented to estimate Schmidt numbers that cannot be calculated from pure fidelity Schmidt witnesses. Pure fidelity Schmidt witnesses can be written as a linear combination of the identity and a projection onto a pure, entangled state. In Fig.~\ref{fig:SDPSchmidt} we computed the first semidefinite programming (SDP) hierarchy. This computation can give a lower bound on the Schmidt number of a mixed state, whereas the second SDP hierarchy proposed in Ref. \cite{Weilenmann} can prove for a state that its Schmidt number cannot be detected with a fidelity witness.

As shown in Fig.~\ref{fig:SDPSchmidt}, bound entangled states are not detectable as entangled by the first SDP hierarchy. On the other hand, the hierarchy detects states as having Schmidt number $3$ beyond those detectable by a pure fidelity Schmidt witness (insets of Fig.~\ref{fig:SDPSchmidt}).

\section{Conclusion}
In summary, we have discussed the entanglement properties 
of a family of highly symmetric, bipartite, mixed quantum 
states in arbitrary dimensions, which can be seen as a generalization of axisymmetric states. We were able to solve
the separability problem for a subfamily, including the 
characterization of bound entanglement, and characterized 
also the dimensionality of entanglement for some cases. 

There are several directions in which our work may be extended.
First, one may study bipartite quantum states with the given symmetry further. Here, it would be desirable to find bound
entangled states far away from the border of separability or 
to determine entanglement monotones, such as the entanglement
of formation for the considered family of states.
In addition, the symmetric bound entangled states in our family, defined in arbitrary dimensions, comprise a promising testbed for finding simpler counterexamples to the Peres conjecture, which stated that bound entangled states cannot display nonlocality. The known counterexamples to date \cite{Vertesi2014,Moroder2014a,Yu2017} are constructed \emph{ad hoc}, and are not defined through symmetries. Finding counterexamples that arise more naturally, as our family of states, could shed more light on whether the original statement by Peres is true except for specially handcrafted examples or it is more generally false.
Finally, one
may consider the multipartite scenario, where states with the
given symmetries can also be defined in a straightforward manner. 
Then, it would be useful to discuss the different classes of
multiparticle entanglement in this scenario.

\section{Acknowledgements}
We thank Mohamed Barakat and Xiao-Dong Yu for discussions. 
This work was supported by the Deutsche Forschungsgemeinschaft (DFG, German Research Foundation, project numbers 447948357 and 440958198), the Sino-German Center for Research Promotion 
(Project M-0294),  ERC (Consolidator Grant 683107/TempoQ),  grant PGC2018-101355-B-I00 funded by MCIN/AEI/ 10.13039/501100011033 and by “ERDF A way of making Europe” and Basque Government grant IT986-16.
M.S.B. acknowledges support from the House of Young Talents 
of the University of Siegen.
G.S. acknowledges support from the Alexander von Humboldt Foundation, from the Spanish Agencia Estatal de Investigaci\'on (project PID2019-107609GB-I00), and from the
Generalitat de Catalunya CIRIT (2017-SGR-
1127).

\onecolumngrid

\bibliography{mybib}

\newpage
\appendix

\section{Some proofs and calculations}

\subsection{Form of the density matrix $\varrho^{\diamond}$} 
\label{sec:FullMatrix}
 The density matrices $\varrho^{\diamond}$ of states that are invariant under these symmetries can be written in the form \\\\
\setcounter{MaxMatrixCols}{30}
\NiceMatrixOptions{code-for-first-row = \scriptstyle,code-for-first-col = \scriptstyle,parallelize-diags=false }
\begin{equation}
\addtocounter{MaxMatrixCols}{5}
\varrho^{\diamond}=
\begin{pNiceMatrix}[baseline=2,last-row,last-col,xdots/line-style={dashed}]
x_{1} 	&0      	& 0    	&      	&  	&\Vdots	&     	&y_{d-1}	&	&	&	& \Vdots	&	&	&y_{2}	&	&\dots 	& y_{1}	&	&	\\
0     	& x_{2} 	& 0    	&      	&  	&   	&   	&	&	&	&	&   	&	&	&	&	&	&	&	&	\\
0     	& 0     	& x_{3}	&      	&  	&   	&   	&	&	&	&	&   	&	&	&	& 	&	&	&	&	\\
        	&       	&        	&\ddots	&  	&   	&      	&	&	&	&	&   	&	&	&	&	&\ddots	&	&	&	\\
        	&       	&        	&	&  x_{d} 	&   	&      	&	&	&	&	&   	&	&	&	&	&	&	&	&	\\
 	&  	&  	& 	&	&   	&     	&     	&     	&      	&	&   	&	&	&	&	& \hdots	&	&	&	\\
 	&  	&  	& 	&      	&   	&x_{d}	&     	&     	&      	&	&   	&	&	&	&	&	&	&	&	\\[-5pt]
 y_{1}	&  	&  	& 	&      	&   \Vdots	&     	&x_{1}	&     	&      	& 	&   \Vdots	&	&	&y_{d-1}	&	& 	&	&	&	\\
  	&  	&    	& 	&      	&   	&     	&     	&x_{2}	&      	&       	&   	&	&	&	&	&	&	&	&	\\[-5pt]
  	&  	&    	& 	&      	&   	&     	&     	&     	&\ddots	&       	&   	&	&	&	&	&\ddots	&	&	&	\\
  	&  	&    	& 	&      	&   	&     	&     	&     	&      	&x_{d-1}	&   	&       	&	&	&	&	&	&	&	\\
  	&  	&    	& 	&      	&   	&     	&     	&     	&      	&       	&   	&	&     	&	&	& \hdots	&	&	&	\\
  	&  	&    	& 	&      	&   	&     	&     	&     	&      	&       	&   	&x_{d-1}	&     	&	&	&	&	&	&	\\
  	&  	&    	& 	&      	&   	&     	&     	&     	&      	&       	&   	&       	&x_{d}	&     	&	&	&	&	&	\\[-5pt]
y_{2}	&  	&    	& 	&      	 & \Vdots	&     	& y_{1}    	&     	&      	&       	&  \Vdots	&       	&     	&x_{1}	&	&	&	&	&	\\
  	&   	&    	& 	&      	&   	&     	&     	&     	&      	&       	&   	&       	&     	&     	&x_{2}	&	&	&	&	\\[-5pt]
 \vdots 	&  	& \ddots   	& 	&      	&   \vdots 	&     	&     	&     	&\ddots      	&       	&   \vdots	&       	&     	&     	&	&\ddots	&	&	&	\\
 y_{d-1} 	&  	&    	& 	&      	&   	&     	&     	&     	&      	&       	&   	&       	&     	&     	&	&	&x_{1}	&	&	\\
	&  	&  	&  	&  	&   	&  	&  	&  	&  	&  	&   	&  	&  	&  	&  	&  	&  	&  	&  	

\CodeAfter
\tikz \draw [dashed] (6.5-|1) -- (6.5-|17) ;
\tikz \draw [dashed] (12.5-|1) -- (12.5-|17) ;
\end{pNiceMatrix}.
\end{equation}

\subsection{Decomposition into separable states for $\varrho_{\text{sep}}$}\label{Sec:Decomp}
 To prove that $\varrho_{\text{sep}}$ is indeed separable we give a decomposition into separable states. First we define $2^d$ states of the following form:
\begin{equation}
  \ket{\varphi_{b}}:=\frac{1}{d} \sum_{k=0 }^ {d-1} (-1)^ {b_k} e^{i\omega k } \ket{k}\quad \forall b\in \lbrace 0,1 \rbrace^d.
\end{equation}
We consider states of the form $ \ket{ \varphi_b}\otimes \ket{\varphi_b^{*}}$ to decompose $ \varrho_{\text{sep}}$ into separable states. 

The following calculation shows the decomposition:
\begin{align}
\nonumber
& \frac{1}{2\pi 2^d} \int \text{d}\omega  \sum_{b\in \lbrace 0,1 \rbrace^d} \ket{ \varphi_b}\otimes \ket{\varphi_b^{*}}\bra{ \varphi_b}\otimes \bra{\varphi_b^{*}}\\ \nonumber
&=\frac{1}{2\pi 2^dd^2} \int \text{d}\omega  \sum_{b\in \lbrace 0,1 \rbrace^d} \sum_{j,k,l,m} (-1)^{b_j+ b_k+b_l+b_m} e^{i \omega (j-k-l+m)}\ket{ j}\otimes \ket{k}\bra{ l}\otimes \bra{m}\\ \nonumber
&=\frac{1}{ 2^dd^2}  \sum_{b\in \lbrace 0,1 \rbrace^d} \sum_{j,k,l,m} (-1)^{b_j+ b_k+b_l+b_m} \delta_{j-k,l-m} \ket{ j}\otimes \ket{k}\bra{ l}\otimes \bra{m}\\ \nonumber
&=  \frac{1}{d^2} \sum_{j,k,l,m} \left[(\delta_{j,k}\delta_{l,m}+\delta_{j,m}\delta_{l,k} )(1-\delta_{j,l}) + \delta_{j,l}\delta_{k,m}\right] \delta_{j-k,l-m} \ket{ j}\otimes \ket{k}\bra{ l}\otimes \bra{m}\\ \nonumber
&=  \frac{1}{d^2} \sum_{j,k,l,m} \left[\delta_{j,k}\delta_{l,m}(1-\delta_{j,m}) +\delta_{j,l}\delta_{k,m}  \right] \delta_{j-k,l-m} \ket{ j}\otimes \ket{k}\bra{ l}\otimes \bra{m}\\ \nonumber
&= \frac{1}{d^2}  \sum_{j,k,l,m} \left[\delta_{j,k}\delta_{l,m}(1-\delta_{j,m}) +\delta_{j,l}\delta_{k,m}  \right]  \ket{ j}\otimes \ket{k}\bra{ l}\otimes \bra{m}\\ 
&=  \frac{1}{d^2} \sum_{j,l\quad j\neq l } \ket{jj}\bra {ll} +\frac{1}{d^2}\sum_{j,k }\ket{jk}\bra{jk} =\varrho_{\text{sep}}.  \label{SuperCalc}
\end{align}

\subsection{Example for an entangled state that is not detected by CCNR}\label{sec:AppPPTEnt}
Consider the state with 
\begin{equation}
    x_1= \frac{1}{80},  \quad x_2=x_3 = \frac{19}{160}. 
\end{equation}
Since $x_3=0<x_1$ this state is entangled according to the above considerations. However calculating the formula \eqref{CCNRZd} yields a value smaller than 1.

\subsection{Determining upper bounds for Schmidt coefficients by looking at pure states with circulant coefficient matrix}
\label{sec:AppSchm}
For the general case in Eq.~\eqref{paramD3} it is difficult to calculate the number of non-vanishing singular values of the coefficient matrix analytically. Although it is possible to obtain the Schmidt rank numerically (uniformly), randomly sampling the pure states will almost certainly lead to a state with full Schmidt rank. 

However if we consider states with $a=b=c$ and $e=f=g$, the coefficient matrix is circulant and one can easily calculate the eigenvalues.  
We get a particular easy example, if we set $\frac{1}{\sqrt{3}}=a=b=c=e=f=g$. For the case $\overline{r}=0 $ we obtain that all coefficients are equal:
\begin{equation}
\sqrt{\frac{z}{3 }}= \sqrt{\frac{1-z}{6 }}
\quad \Rightarrow z = \frac{1-z}{2} 
 \quad  \Rightarrow  z= \frac{1}{3} .
\end{equation}
Since all entries of the coefficient matrix are equal the matrix has only one eigenvalue and therefore Schmidt number one.
Further the state is indeed twirled to $\varrho_{\text{sep}}$, which was shown to be separable above. 
Another state we can construct has the parameters $\frac{-1}{\sqrt{3}}=a=b=c=e=f=g$. Since the coefficient matrix is circulant, one of the three eigenvalues is simply the sum over all different entries of the matrix. To get an upper bound of the Schmidt number of the state $\sigma$, we set this particular eigenvalue to zero for $\overline{r}=0$:
\begin{equation}
0=\sqrt{\frac{z}{3 }} - 2 \sqrt{\frac{1-z}{6 }} \quad \Rightarrow  z= 2 (1-z ) \quad \Rightarrow z = \frac{ 2}{3}. 
\end{equation} 
This result is in agreement with the first Schmidt number witness $W_2$ in Eq. \eqref{eq:firstWitness} that we applied, because the Schmidt witness is tangent to the state.
Another state we can investigate is at $\overline{r}=1 $. We again set the same eigenvalue equal to zero and keep in mind that the coefficients with $\sqrt{1-r}$ vanish:
\begin{equation}
0=\sqrt{\frac{z}{3 }} - \sqrt{2} \sqrt{\frac{1-z}{6 }} \quad \Rightarrow  z=  (1-z ) \quad \Rightarrow z = \frac{ 1}{2}.
\end{equation}

For the $\overline{r}$-values in between we still can look at a circulant coefficient matrix and set one eigenvalue equal to zero,

\begin{align} \nonumber
&0=\sqrt{\frac{z}{3}}- \sqrt{1-z } (\sqrt{\frac{1+\overline{r}}{6}}+\sqrt{\frac{1-\overline{r}}{6}}) \\\nonumber
&\Leftrightarrow \sqrt{x_1}= \sqrt{x_2}+ \sqrt{x_3}\\\nonumber
&\Rightarrow x_1 = x_2+ x_3 + 2\sqrt{x_2 x_3 }\\ \nonumber
&\Rightarrow 2x_1 = \frac{1}{3} + 2\sqrt{x_2 x_3 }\\ 
& \Rightarrow  x_1= \frac{1}{6}\left( 1- 3 x_2 \pm \sqrt{ 3}\sqrt{2x_2 - 9 x_2^2} \right), \label{CurveYB}
\end{align}
where we used the normalization condition $x_1+ x_2 +x_3= 1/3$. 
\\\\

\subsection{States with Schmidt-rank$ < d$} \label{sec:NonFullSR}
We saw that some states with Schmidt rank smaller than $d$ are characterized by

\begin{align}
&\sqrt{x_1}=\sum _{k=1 }^{d-1} \sqrt{x_{k+1}}. \\
\intertext{ With the normalization condition $ \frac{1}{d}= \sum_{k=1}^{d}x_k$ we get:}
&\sqrt{x_1}= \sqrt{\frac{1}{d}- x_1 -\sum_{m=3}^d x_m  }+ \sum _{k=3 }^{d} \sqrt{x_{k}}. 
\end{align}

To solve this for $x_1$ one has to essentially solve some quadratic equation. The solution is
\begin{equation}
  x_1= \frac{1}{2}   \left(  \frac{1}{d} - \sum_{k=3}^d  x_k  - \left(\sum_{k=3}^d \sqrt{x_k} \right) \sqrt{4 \left(\frac{1}{d} - \sum_{m=3}^d x_m \right)+\left(\sum_{m=3}^d \sqrt{x_m}\right)^2  }\right),
\end{equation}
which is the generalization of \eqref{CurveYB} to higher dimensions.

\end{document}